\documentclass[aps,twocolumn,pra,amsmath,amssymb,aps]{revtex4-1}
\usepackage{graphicx}
\usepackage{dcolumn}
\usepackage{bm}
\usepackage{graphicx,lastpage}
\usepackage{graphics}
\usepackage{amsmath}
\usepackage{amssymb}
\usepackage{upgreek}
\usepackage{color}
\usepackage{float}
\usepackage{soul}
\usepackage{ragged2e}
\usepackage{subfig,caption,subcaption}
\usepackage{censor}
\usepackage{enumerate}
\usepackage{textcomp}
\usepackage{braket}
\usepackage{placeins} 
\usepackage{ulem}
\usepackage{xr}

\begin{document}

\title{Thermodynamics of a biophotomimetic nonreciprocal quantum battery}

\author{Trishna Kalita}
\author{Manash Jyoti Sarmah}
\author{Himangshu Prabal Goswami}
\email{hpg@gauhati.ac.in}
\affiliation{QuAinT Research Group, Department of Chemistry, Gauhati University, Jalukbari, Guwahati-781014, Assam, India}
\date{\today}

\begin{abstract}
We propose a theoretical model of a fully functional nonreciprocal quantum battery inspired by the architecture of bacterial light-harvesting complexes. We assign functional roles to collective quantum optical subradiant and superradiant states and introduce a unimodal cavity to assist storage.  The transition rates are obtained from an effective non-Hermitian Hamiltonian, tailored to the battery geometry which are fed into a master equation to unravel the time evolution. We investigate the complete thermodynamic performance including storage, leakage, ergotropy, work extraction, flux, and power. We observe optimization at different ring sizes, each peaking at its specific energetic function. Strong coupling between the ring and central system enhances the battery's ability to store energy but reduces the ability of power output. The ergotropy exceeds capacity and approaches it linearly with increasing system size, with an optimal small-size regime that disappears under strong coupling.

\end{abstract}

\maketitle

\section{Introduction}

Finite-dimensional quantum systems capable of storing and releasing energy in ways unnatural to classical devices  make the broad field of quantum batteries (QB) \cite{Alicki2013, Andolina2019PRL, binder2015quantacell, campaioli2019quantum, bhattacharjee2021quantum}. Theoretical studies have shown that collective effects such as entanglement~\cite{Ferraro2018} and cooperative charging~\cite{Campaioli2017} can enhance performance beyond standard limits, enabling superextensive scaling of stored energy or charging power. For instance, in the Dicke-model QB, global operations can induce entanglement between a number of identical cells, producing enhanced charging-power scaling that surpasses the natural scaling of independent cells \cite{binder2015quantacell, campaioli2017enhancing}.  Realistic devices, on the other hand, are open quantum systems subject to dissipation, dephasing, and leakage, which limit both storage capacity and charging speed~\cite{mitchison2019charging,santos2019quantum,rossini2019quantum,barra2019dissipative}. This has motivated the search for architectures where cooperative advantages survive in the steady state, including dissipative QBs~\cite{rossini2019quantum}, and those driven by central chargers~\cite{han2023steady,chen2022optimal}. QBs based on spin chains \cite{campaioli2017enhancing}, cavity-QED setups \cite{binder2015quantacell,ferreira2022ergotropy}, and excitonic aggregates \cite{rossini2019quantum} have also been proposed.

The quest for the design of efficient energy storage,
driven
in parallel to theories of fundamental understanding, has drawn numerous inspiration from natural light-harvesting complexes (LHCs). It is now well accepted that structured dipolar arrangements of biosystems  can uplift excitonic transport efficiency under tuned conditions~\cite{Scholes2011,Blankenship2014} of symmetry, delocalized excitations, and engineered dissipation pathways. Such structural knowledge allow identification of control parameters to funnel energy toward reaction centers with minimal losses~\cite{van2000photosynthetic, Scholes2011}. The purple bacterial LHC  comprising of a nearly complete ring of bacteriochlorophyll dipoles surrounding a central reaction center (RC) is a canonical example of structural optimization for energy capture and conversion~\cite{Hu1997architecture,ritz2001long}. Such LHC RC complexes exhibit long and short lived states usually called the sub and superradiant states. Also called bright and dark pairs, these naturally emerge when such complexes are modeled as dipolar rings geometries via symmetry-induced constructive and destructive interference of radiative amplitudes \cite{monshouwer1997superradiance,cho2005exciton}. The subradiant or dark modes are long-lived, making these ideal for storage.

The performance of a quantum battery is also characterized by how much of the stored energy can be converted into useful work apart from the standard charging power and leakage channels. In this context, ergotropy  \cite{Allahverdyan2004} and battery capacity \cite{yang2023battery} are two exotic figures of merit. Ergotropy distinguishes stored energy from extractable work \cite{Allahverdyan2004,Perarnau2015,Francica2020} and is particularly relevant  where dissipation and decoherence can degrade the recoverable work content even when the stored energy remains appreciable \cite{Cakmak2020,Tirone2023}. Complementary to ergotropy, the battery capacity \cite{yang2023battery, Zhang2024MeasurementEnhancedCapacity} characterizes the potential of an energy-storing quantum system to store and supply energy. Together, power, ergotropy, and battery capacity provide a quantitative framework for assessing the performance of QBs. These allow a construct on capturing  fast energy  deposition, work extraction, and understanding work delivery through system parameters.

Building forward from known principles, we propose a quantum battery that structurally mimics a biological  LHC RC and functions as a quantum battery. Envisioning it as a set of identical dipoles arranged in a ring around a central system mimics the core architecture of the LHC RC complex. The superradiant states of such a ring system interacting with a central moeity allows rapid energy intake \cite{moreno2019subradiance}, thereby acting as a fast charger.  Although, this division of functionality between bright and dark states has been exploited in quantum light-harvesting studies \cite{giusteri2016light}, we shall adapt it here to explore the ergotropy and battery capacity. Auxilliary states are also included so as to mimic higher vibronic manifolds or charge-transfer intermediates~\cite{novoderezhkin2011physical,huh2014exciton}, features known in biological systems to facilitate energy regulation \cite{Creatore2013}. Analogous multi-level QB architectures have been shown to enhance robustness to noise and improve charging efficiency~\cite{zhang2019enhanced,santos2019quantum}.  To gain a control over reversible energy exchange channels, we also introduce a cavity to the battery's storage subspace. Through the interaction with a charging bath, which is a friendly experimental platform \cite{PhysRevA.108.052213} and can also enhance charging power \cite{lu2025topological}, the overall QB becomes inherently non-reciprocal \cite{ahmadi2024nonreciprocal}.  

The theoretical treatment of the proposed quantum battery is based on a master equation \cite{timm2008tunneling}, where the rates are not phenomenological but derived from the diagonalization of an effective non-Hermitian Hamiltonian  \cite{Novotny2012, Pozas2016comprehensive, Moiseyev2011non, lee2014heralded} which fully accounts for the geometry-dependent coupling between the involved quantum states and the electromagnetic environment. This approach naturally incorporates retardation effects, collective Lamb shifts, and cooperative decay channels. Such a formalism has also been widely employed in studies of superradiance and subradiance 
in pigment–protein complexes \cite{caruso2009highly,wu2010efficient}, where geometry-dependent dipolar interactions were shown to bridge microscopic excitonic dynamics with open-system thermodynamics.  
The paper is organised as follows. In Sec. II we introduce the ring–central-system geometry and  discuss the bright and dark modes from an effective non-Hermitian description. In Sec. III, we introduce the overall Hamiltonian and formulate the open-system master equation. In Sec. IV we define the relevant thermodynamic quantities, charging energy, leakage energy, ergotropy, capacity, flux, work, and power. We present our  results and relevant discussions on the varying trends with ring size and system coupling in Sec. V  after which we conclude.

\section{Quantum Battery Model}

\begin{figure}[h]
\centering
\includegraphics[width=1\linewidth]{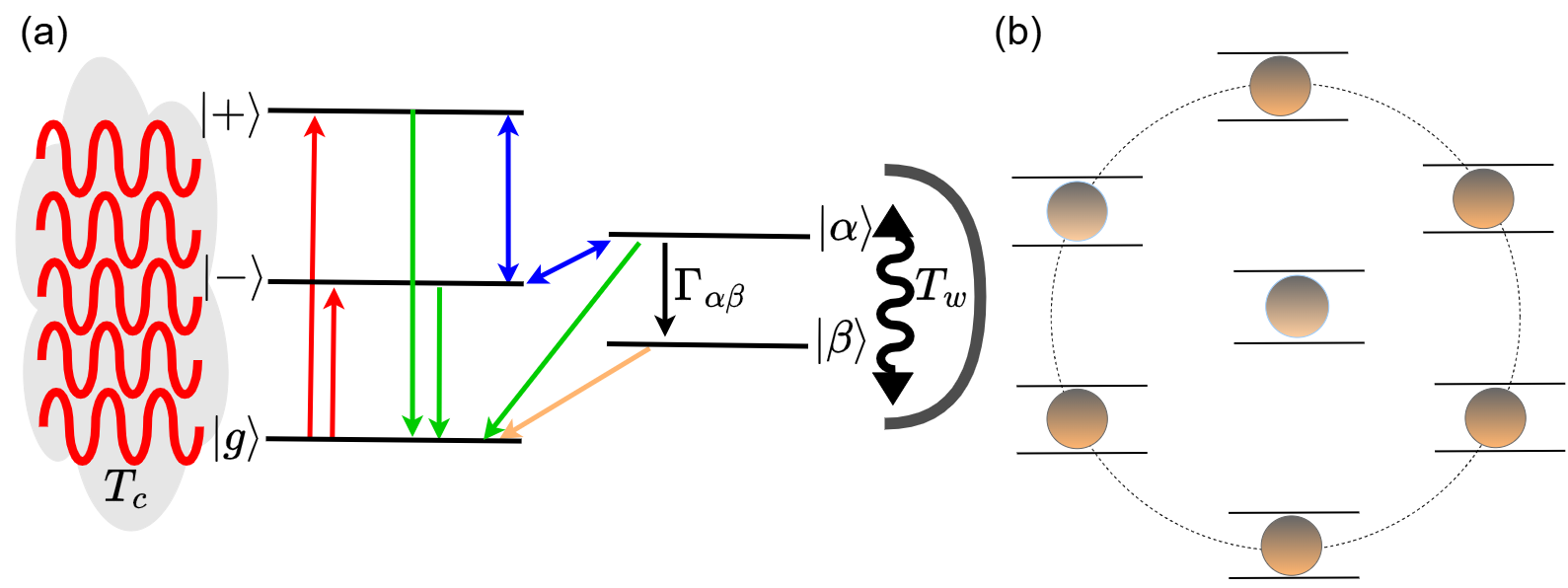}
\caption{
(a) Schematic representation of the quantum battery. The manifold $\{|+\rangle, |-\rangle, |g\rangle\}$ is coupled to a charging reservoir at temperature $T_c$, while the states $\{|\alpha\rangle, |\beta\rangle\}$ are coupled to a unimodal cavity at a temperature $T_w$. The red arrows denote charging, the green arrows denote leakage, and the blue arrows denote storage.
(b) Light-harvesting-inspired geometry of $N_R$ number of  two-level systems (whose decay rate is $\Gamma$) arranged in the form of an equally spaced symmetric ring and coupled to a central two-level system (detuning $\Delta$, decay rate $\Gamma_0$). The collective eigenstates of the coupled system give rise to bright (superradiant, $|+\rangle$) and dark (subradiant, $|-\rangle$) states and can be estimated from Eq. (2).} 
\label{fig:ring_qbat_DA}
\end{figure}

 The architecture we propose is biomimetically motivated by natural light-harvesting (LH) complexes.  For example, the model is a  mimic the cooperative LHC RC architecture found in purple bacteria \cite{}. In these biological assemblies, the LHI complex consists of bacteriochlorophyll-protein units arranged in a ring-like geometry, which absorb excitation energy from the surrounding LHII complexes and subsequently funnel it toward a central reaction center. It is akin to a ring-central-emitter construction \cite{moreno2019subradiance} that translates the biological design principle into a minimal quantum battery setting. The ring with the central system produces collective bright and dark modes.  Overall, the battery has five quantum states that are asymmetrically coupled to a bosonic  reservoir and a unimodal cavity, as depicted schematically in Fig.~\ref{fig:ring_qbat_DA}(a). The relevant states are denoted by \(|g\rangle\), \(|+\rangle\), \(|-\rangle\), \(|\alpha\rangle\), and \(|\beta\rangle\), where \(|g\rangle\) is the ground state, \(|+\rangle\) and \(|-\rangle\) are excited states, and \(|\alpha\rangle\) and \(|\beta\rangle\) are auxiliary states. The reservoir allows directional energy flow due to multiple incoherent baths \cite{}.
Reservoir-induced transitions $\ket{g}\leftrightarrow\ket{+}$ and $\ket{g}\leftrightarrow\ket{-}$ inject energy into the system and therefore define the charging channels.   By contrast, the $\ket{-}$ is long-lived, making it a natural choice for storage state. To realize a realistic cycle of energy flow, the dark state $\ket{-}$ is coupled to an acceptor-like state $\ket{\alpha}$, which is further connected to a lower-energy state $\ket{\beta}$. The coupling between $\ket{-}$ and $\ket{\alpha}$ allows stored excitation to be transferred into a sector that further protects energy before release.  The state $\ket{\beta}$, which is populated via cavity-induced transitions from $\ket{\alpha}$ and decays to $\ket{g}$, serves as the discharge channel. Since the two auxiliary states $\ket{\alpha}$ and $\ket{\beta}$ mimic downstream acceptor and release channels, in this sense, these mimic the functional logic of photosynthetic energy capture, transfer, and temporary retention. We accommodate the cavity into the design so as to control the storage subspace. Further, the cavity also serves as a control unit to introduce non-reciprocity into the battery operation.

The incoherent charging operations and stochastic coupling with the reservoirs and cavity inherently introduce energy and particle fluctuations that effect the populations leading to active and passive states, naturally allowing one to explore the concept of ergotropy and battery capacitance.
Accordingly, the model functions as a quantum battery because it contains all essential operational ingredients: an uncharged reference state $\ket{g}$, charging pathways from the reservoir, an internal storage sector formed mainly by $\ket{-}$ and $\ket{\alpha}$, leakage channels through the bright manifold, and a controlled discharge route through $\ket{\beta}$.  The system can relax to $\ket{g}$ and thus complete the charging-storage-leakage-discharge cycle.  Unlike Dicke-type batteries, in which all cells couple symmetrically to a common charger \cite{Ferraro2018}, the present model exploits geometry-induced superradiant and subradiant manifolds to achieve a natural division between rapid charging and long-lived storage.

The $\ket{+}$ and $\ket{-}$ states can be explicitly formulated from a light-harvesting-inspired submodel consisting of $N_R$ identical two-level systems arranged on a symmetric ring and coupled to a central two-level system, as shown in Fig.~\ref{fig:ring_qbat_DA}(b) \cite{hu1997pigment,ritz2002quantum,olaya2008efficiency}. Within the single-excitation manifold, this coupled ring-central architecture supports collective hybridized states \cite{holzinger2024harnessing}, where it has been established that one can restrict attention to the dominant symmetric ring mode by working with a effective non-Hermitian Hamiltonian,
\begin{align}
\hat{H}_{\mathrm{eff}}
&=
\left(\Delta-i\frac{\Gamma_0}{2}\right)\hat{\sigma}_d^\dagger\hat{\sigma}_d
+
\left(\tilde{J}_m-i\frac{\tilde{\Gamma}_m}{2}\right)\hat{S}_m^\dagger\hat{S}_m
\nonumber\\
&\quad
+\sqrt{N_R}\left(J_d-i\frac{\Gamma_d}{2}\right)
\left(\hat{S}_m^\dagger\hat{\sigma}_d+\hat{S}_m\hat{\sigma}_d^\dagger\right),
\label{eq:Heff_model_rearranged}
\end{align}
where $\Delta$ and $\Gamma_0$ are the detuning and decay rate of the central system, $\hat{\sigma}_d$ describes the central two-level system. and $\hat{S}_m$ denotes the symmetric collective ring mode. The quantities $J_d$ and $\Gamma_d$ describe coherent and dissipative  coupling between the ring and central system. $\tilde{J}_m$ and $\tilde{\Gamma}_m$ are the collective energy shift and radiative decay rate of the isolated symmetric ring mode.

When we assume identical pairwise coherent and dissipative couplings within the isolated ring (without the central system), then $\Omega_{ij}=\Omega$ and $\Gamma_{ij}=\Gamma$.  The collective modes' eigenvalues of the isolated ring are
\begin{align}
\zeta_m
=
\frac{1}{N_R}\sum_{j,\ell}^{N_R}
\exp\!\left[\frac{2\pi i m}{N_R}(\ell-j)\right]
\left(\Omega-i\frac{\Gamma}{2}\right),
\end{align}
with $m=N_R/2$ for even $N_R$ and $m=(N_R-1)/2$ for odd $N_R$ \cite{moreno2019subradiance}. The corresponding collective shift and decay are
$
\tilde{J}_m=\mathrm{Re}(\zeta_m),
\qquad
\tilde{\Gamma}_m=-2\,\mathrm{Im}(\zeta_m).
$
Diagonalization of $\hat{H}_{\mathrm{eff}}$ yields two complex eigenvalues,
\begin{align}
\lambda_{\pm}(\Delta)
&=
\frac{1}{2}
\left(
\tilde{J}_m-\Delta
-\frac{i}{2}(\tilde{\Gamma}_m+\Gamma_0)
\right)
\nonumber\\
&\mp
\sqrt{
\left(
\tilde{J}_m+\Delta
-\frac{i}{2}(\tilde{\Gamma}_m-\Gamma_0)
\right)^2
+
4N_R\left(J_d-\frac{i}{2}\Gamma_d\right)^2
},
\label{eq:eigenvalues_split_rearranged}
\end{align}
whose real parts define the hybridized energies
$
E_+=\mathrm{Re}[\lambda_-],
\qquad
E_-=\mathrm{Re}[\lambda_+].
$
The imaginary parts of the eigenvalues determine the effective lifetimes of the $\ket{+}$ and $\ket{-}$ states. $
\Gamma_+=-2\,\mathrm{Im}(\lambda_-),
\qquad
\Gamma_-=-2\,\mathrm{Im}(\lambda_+).
$
The scaled decay rates of these two states are shown in Fig.~\ref{Gamma_+,-} (a) and (b).
The state $\ket{+}$ is a superradiant or bright state which exhibits an enhanced decay rate in comparison to the state $\ket{-}$ which has a  reduced decay rate due to destructive interference. It is hence, a subradiant or dark state.  Since $\ket{+}$ possesses a comparatively larger decay rate, it also serves as a lossy intermediate. A part of the injected energy can leak back to the environment through it. 

A similar battery based on ring arrays has been recently proposed where a charger atom is pumped by an incoherent thermal source \cite{PhysRevA.111.032221}. The work is mainly a many-body quantum battery made of a ring of atoms with collective charging, while the model described here is a single multilevel quantum system charged directly by a thermal bath. The referenced work treats the central atom as an external charger that incoherently pumps a ring of two-level atoms, whereas in the current work,  the central system is not an external charging unit but an intrinsic part of the quantum system itself.  Moreover, we include a unimodal cavity that selectively couples two separate    transitions allowing the stored excitation to be converted into a controllable output, which is absent in the referenced ring-array model. Consequently, the model described in this work opens the possibility of studying cavity-assisted energy extraction, and tunable energy flow between thermal pumping and work extraction, offering an alternative platform beyond the collective charging paradigm.
\begin{figure}[h]
\centering
\includegraphics[width=1\linewidth]{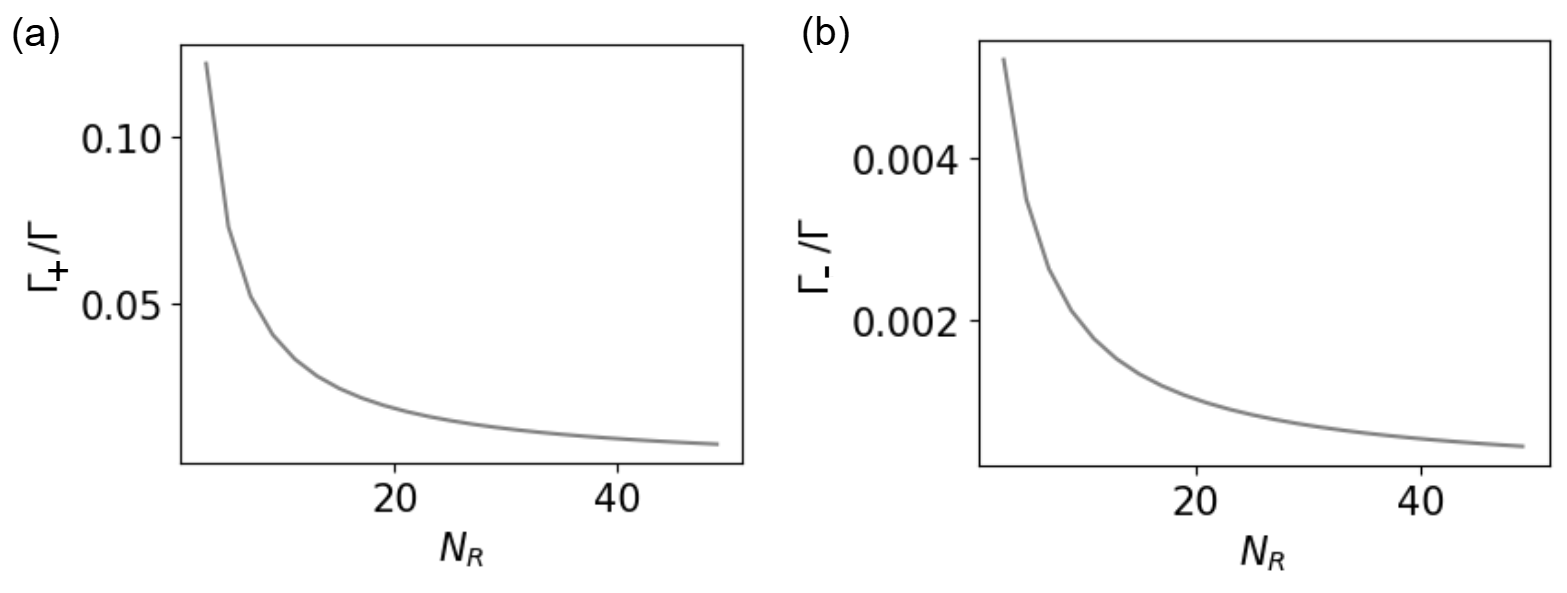}
\caption{Scaled decay rates of the bright state, $\Gamma_{+}/\Gamma$, and the dark state, $\Gamma_{-}/\Gamma$, evaluated using Eq.~(\ref{eq:eigenvalues_split_rearranged}). Parameters used are $\Omega=0.5$, $\Gamma=0.8$, $\Delta=0.5$, $J_d=2$, $\Gamma_d=0.0002$, and $\Gamma_0=0.5\tilde{\Gamma}_m$. We have set $\hbar,k_B\to 1$ so that natural units are employed in the numerics.}
\label{Gamma_+,-}
\end{figure}

\section{ Methodology}
We are now in a position to quantitatively specify the battery model. 
The quantum battery model in our work now represents a five-level quantum junction coupled to an external reservoir and a unimodal cavity. Three states,  $\ket{+}$, $\ket{-}$, and ground $\ket{g}$ are coupled to a source or charging reservoir ($T_c$). Two acceptor states $\ket{\alpha}$ and $\ket{\beta}$, are coupled to  a unimodal cavity ($T_w$) illustrated in Fig. (~\ref{fig:ring_qbat_DA} b). Energy exchange is possible between $\ket{-}$ and $\ket{\alpha}$ as well as $\ket{\beta}$ and $\ket{g}$. The total Hamiltonian of the system is given by:
\begin{equation}
    \hat{H} = \hat{H}_{\text{q}} + \hat{H}_{\text{E}}+\hat H_w + \hat{V}_{\text{I}},
\end{equation}
where \( \hat{H}_{\text{q}} \) is the bare system's Hamiltonian. \( \hat{H}_{\text{E}} \) describes the charging source's bosonic environment, and $\hat H_w$ is the quantum cavity's Hamiltonian. \( \hat{V}_{\text{I}} \) captures the system-surrounding interaction.
The central quantum system's Hamiltonian is
\begin{align}
    \hat{H}_{\text{q}} 
    &= \sum_{i} \epsilon_{i} \ket{i}\bra{i} \nonumber \\
    &\quad + \left(  
        V_{-\alpha} \ket{-}\bra{\alpha} 
        + V_{\beta g} \ket{\beta}\bra{g} 
        + V_{\alpha g} \ket{\alpha}\bra{g} 
        + \text{H.c.} 
    \right).
\end{align}
where $\quad \ket{i} \in \{\ket{g}, \ket{+}, \ket{-}, \ket{\alpha}, \ket{\beta} \}$. The basis of the Hamiltonian is arranged in increasing energy, i.e $\epsilon_i\le \epsilon_{i+1}$. The environment Hamiltonian is
\begin{equation}
    \hat{H}_{\text{E}} =  \sum_{k} \hbar\omega_{k} \hat{b}_{k}^\dag \hat{b}_{k}
\end{equation}
where \( \hat{b}_{k}^\dag \) and \( \hat{b}_{k} \) are the bosonic creation and annihilation operators for mode \( k \) of the charging source.
The quantum cavity's Hamiltonian is
\begin{equation}
    \hat H_w=\hbar\omega_w\hat a^\dag\hat a
\end{equation}
 $\hat a (\hat a^\dag) $ being the cavity's boson annihilation (creation) operator at its frequency $\omega_w$.
The system-surrounding interaction Hamiltonian consists of couplings induced by the charging source and unimodal cavity,
$
    \hat{V}_{\text{I}} = \hat{V}_c + \hat{V}_w
$. The charging reservoir induces transitions among \( \ket{g}, \ket{+}, \ket{-} \):
\begin{align}
    \hat{V}_c &= \sum_{k} \big[ 
    g_{k}^{g+} \ket{g}\bra{+}
    + g_{k}^{g-} \ket{g}\bra{-}
    + g_{k}^{+-} \ket{+}\bra{-}
    \big]\nonumber\\
    &\times(\hat{b}_{k}^{\dagger} + \hat{b}_{k}) + \text{H.c.}
\end{align}
while the unimodal cavity induces transitions between \( \ket{\alpha} \) and \( \ket{\beta} \):
\begin{equation}
    \hat{V}_w =  g^{\alpha\beta} \ket{\alpha}\bra{\beta} (\hat a^\dag + \hat a) + \text{H.c.}
\end{equation}
Note that, such unimodal cavities have been used as work modes in quantum heat engines.\cite{sarmah2024noise}

To derive the reduced dynamics of the five-level system under the action of the reservoir and cavity, we move to the interaction picture with respect to the free Hamiltonian \( \hat{H}_0 = \hat{H}_q + \hat{H}_E +\hat H_w\), so that any operator transforms as, $\tilde O(t)=exp(i\hat H_0 t)\hat O\exp(-i\hat H_0t)$, the equation of motion for the total density matrix becomes
\begin{equation}
    \frac{d}{dt} \tilde{\rho}(t) = -i [\tilde{V}(t), \tilde{\rho}_{\text{tot}}(t)].
\end{equation}
Formally integrating this equation and expanding to second order in the interaction strength yields
\begin{align}
    \frac{d}{dt} \tilde{\rho}(t) &= - \int_0^t \mathrm{Tr}_E \left[ \tilde{V}(t), [\tilde{V}(t'), \tilde{\rho}_{\text{tot}}(t')] \right] dt'.
\end{align}
Where $\tilde{\rho}(t)$ is the reduced density matrix of the five-level system and $\mathrm{Tr}_E $ denotes tracing over the external degrees of freedom. We then apply the Born approximation, \( \tilde{\rho}_{\text{tot}}(t') \approx \tilde{\rho}(t') \otimes \rho_E \), where \( \rho_E \) denotes the state of the reservoir and cavity modes external to the five-level system. Under the Markov approximation, we take \( \tilde{\rho}(t') \approx \tilde{\rho}(t) \) and extend the upper limit to infinity, which gives
\begin{equation}
    \frac{d}{dt} \tilde{\rho}(t) = - \int_0^{\infty} \mathrm{Tr}_E \left[ \tilde{V}(t), [\tilde{V}(t - \tau), \tilde{\rho}(t) \otimes \rho_E] \right] d\tau.
\end{equation}
To evaluate this expression, we decompose the system operators into transition-frequency components, \( \hat{M}_b(t) = \sum_{\omega} \hat{A}_b(\omega) e^{-i \omega t} \), where \( \hat{A}_b(\omega) = \sum_{\epsilon_j - \epsilon_i = \omega} \ket{i} \bra{i} \hat{M}_b \ket{j} \bra{j} \) projects onto system transitions of frequency \( \omega \). Similarly, we define the environment correlation functions,
$
    C_b(\tau) = \langle \hat{B}_b(\tau) \hat{B}_b(0) \rangle.
$
Substituting and tracing over the environment and employing the secular approximation to discard the fast-oscillating terms between non-degenerate transitions, the master equation reduces to a Lindblad master equation. We would like to point out that the effect of quantum coherences on the thermodynamics of such open and finite quantum systems can be quite nontrivial and we have chosen to work with the secular regime to keep the physics relatively simple \cite{PhysRevA.110.032213, thakuria2025coherence, PhysRevA.110.052214}.  When projecting this equation onto the populations $\rho_{ii} = \langle i | \rho | i \rangle$, and defining the population vector \( \rho = \{ \rho_+, \rho_-, \rho_{\alpha}, \rho_{\beta}, \rho_G \} \) we obtain $
    \dot\rho = \mathcal{L} \rho,
$
where \( \mathcal{L} \) is the rate matrix superoperator capturing transitions among levels given by
\begin{widetext}
\begin{equation}
\resizebox{\linewidth}{!}{
$\mathcal{L} =
\begin{bmatrix}
-(k_{+-} + k_{+g}) & k_{-+} & 0 & 0 & k_{g+} \\
k_{+-} & -(k_{-+} + k_{-\alpha} + k_{-g}) & k_{\alpha -} & 0 & k_{g-} \\
0 & k_{-\alpha} & -(k_{\alpha -} + k_{\alpha\beta} + k_{\alpha g}) & 0 & 0 \\
0 & 0 & k_{\alpha\beta} & -k_{\beta g} & k_{g\beta} \\
k_{+g} & k_{-g} & k_{\alpha g} & k_{\beta g} & -(k_{g+} + k_{g-} + k_{g\beta})
\end{bmatrix},$
}
\end{equation}
\end{widetext}
where the transition rates \(k_{ij}\) are 
$
k_{ij} = \Gamma_{ji} [ n_{x}(\omega_{ji}) + 1 ], 
k_{ij} =  \Gamma_{ij} n_{x} (\omega_{ij}), x = c,w$.
The explicit forms of the  rates are taken as
$k_{+-} = \Gamma_{+-}[n_c(\omega_{+-})+1]$, 
$k_{-\alpha} = \Gamma_{-\alpha}[n_w(\omega_{-\alpha })+1]$, 
$k_{+g} = \Gamma_{+g}[n_c(\omega_{+g})+1]$, 
$k_{\alpha -} = \Gamma_{\alpha-} n_w \omega_{-\alpha }$, 
$k_{-g} = \Gamma_{-g}[n_c(\omega_{-g})+1]$, 
$k_{\alpha g} = \Gamma_{\alpha g} n_w \omega_{\alpha g}$, 
$k_{g+} = \Gamma_{+g} n_c \omega_{+g}$, 
$k_{\alpha \beta} = \Gamma n_w \omega_{\alpha \beta}$, 
$k_{g-} = \Gamma_{g-} n_c \omega_{g-}$, 
$k_{\beta g} = \Gamma_{\beta g} n_w \omega_{\beta g}$, 
$k_{-+} = \Gamma_{-+} n_c \omega_{-+}$, 
and $k_{g\beta} = \Gamma_{g\beta }[n_w(\omega_{g\beta })+1]$. $
n_{x,ij} = (\exp\{\Delta \omega_{ji}/k_b T_{x}\} - 1)^{-1}, x = c,w
$
is the Bose-Einstein distribution. The cavity temperature $T_{w}$ is treated as a fictitious temperature. $\Gamma_{ij}$ are the effective decay rates between the five quantum states. The values of these decay terms are estimated from the imaginary parts of $\lambda_{\pm}$ of the ring with central system as per the following relations. 
$
\Gamma_{+-}=\Gamma_{+g}=0.5\,\Gamma_+,
\qquad
\Gamma_{-+}=\Gamma_{-g}=0.35\,\Gamma_-,
$
$
\Gamma=1.0\times10^{-6},
\qquad
\Gamma_{-\alpha}=0.3\,\Gamma_-,
$
$
\Gamma_{\alpha g}=0.25\,\Gamma,
\qquad
\Gamma_{\alpha-}=0.1\,\Gamma_{-\alpha},
\qquad
\Gamma_{\beta g}=0.5.
$
The $\Gamma_{+}$ is divided equally between the $+\to-$ and $+\to g$ channels, giving
$
\Gamma_{+-}=\Gamma_{+g}=0.5\,\Gamma_{+}.
$
In the same way, $\Gamma_{-}$ is resolved into the three decay channels $-\to+$, $-\to g$, and $-\to \alpha$ with branching fractions $0.35$, $0.35$, and $0.3$, respectively, so that their sum reproduces the full width $\Gamma_{-}$. The remaining prefactors of the rates are chosen as effective phenomenological parameters to preserve a physically reasonable hierarchy among the transition channels, thereby allowing the interplay between charging, storage, and leakage to govern the optimization of the thermodynamic quantities. Note that, the initial quantities $\Omega, \Gamma$ and $\Gamma_0$ can be obtained by simulating an actual arrangement of dipoles in a coordinate system and solving the dyadic Green's tensor \cite{moreno2019subradiance}. We, however treat these as parameters in our calculations.

The time evolution of the five states are obtained by solving the above quantum master equation subject to probability conservation,
$
\rho_{++}+\rho_{--}+\rho_{\alpha\alpha}+\rho_{\beta\beta}+\rho_{gg}=1.
$
At steady state,
$
\dot{\rho}_{ii}=0,
$
and the stationary populations
$
\rho_{++},\quad \rho_{--},\quad \rho_{\alpha\alpha},\quad \rho_{\beta\beta},\quad \rho_{gg}
$
are obtained. These steady-state populations form the basis for analyzing the thermodynamic performance of the battery.

\section{THERMODYNAMIC QUANTITIES}
Firstly we define the charging, storage and leakage Hamiltonians given respectively by,
\begin{align}
\hat H_c &= \sum_{i = g, +,  -} \epsilon_{i} \ket{i}\bra{i}\\
\hat H_s &= \sum_{i = +,-, \alpha} \epsilon_{i} \ket{i}\bra{i}\\
\hat H_L &= \sum_{i =+,-, \alpha, g} \epsilon_{i} \ket{i}\bra{i}. 
\end{align}
The expectation values of these Hamiltonians with respect to the battery's reduced density matrix, $\langle \hat H_c\rangle$, $\langle \hat H_s\rangle$ and $\langle \hat H_L\rangle$ are the charging, storage and leakage energies respectively. One can also define three additionally relevant quantities, the maximum possible charging ($\langle \hat H_c^+\rangle$) and minimum available storage ($\langle \hat H_s^+\rangle$) along with the minimum unavoidable leakage ($\langle \hat H_L^+\rangle$). These three quantities are the expectation values of the above three Hamiltonians with respect to the passive state ($\rho^+$) of the battery and hence are the passive counterparts. These passive states represent the closest states to $\rho$ that cannot yield any work under unitary transformations. For the passive state, the density matrix's ($\rho$) basis is arranged $\rho_{k}\ge \rho_{k+1}$, i.e  descending population provided the total battery Hamiltonian's basis is arranged as $\epsilon_k \le \epsilon_{k+1}$. Note that, throughout the manuscript, expectation values denoted with the superscript $+$ shall represent traces with respect to a passive state, $\rho^+$. Likewise, another density matrix ($\rho^-$) where the population is reordered in ascending basis ($\rho_{k}\le \rho_{k+1}$) is ususally called the anti-passive, antiergotropic, or inverted passive state. It represents  the most excited or the work-rich macrostate configuration. Although, the state is the most energetically unfavorable arrangement compatible with the same set of populations, it tells us how much the system’s internal population structure can contribute to storing energy,  without changing the populations, but by reordering \cite{yang2023battery}. The density matrices, $\rho^-$ and $\rho^+$  represent the most excited and most relaxed population arrangements compatible with the same spectrum of populations, and they quantify the energy span over which the quantum battery can store reversible energetic tension.

Another quantity of interest in quantum batteries is the  concept of ergotropy, which quantifies the maximum work that can be extracted from a quantum system via unitary operations\cite{allahverdyan2004maximal, barra2019dissipative}. The passive states play the determining role in allotting energy availability\cite{sparaciari2017energetic}. Specifically, ergotropy measures the difference between the system's energy in its actual (or active) state and that of its associated passive state. 
\begin{equation}
    {\cal E} = \langle \hat H_q\rangle - \langle \hat H^+_q\rangle.
\end{equation}
In our system, the combined action of the charging reservoir and cavity-driven transfer brings the battery to a nonequilibrium steady state rather than an equilibrium Gibbs state. This can reorder the populations and make the steady state active, so the ergotropy measures how much of the stored excitation can be extracted as useful work. So,
a finite ergotropy is a potential for work extraction.
In analogy to the concept of ergotropy, we also consider the capacity of the quantum battery, which has been introduced as a measure of a battery's energy storage capability \cite{yang2023battery,Zhang2024MeasurementEnhancedCapacity}. The quantity relates the stored energy and ergotropy to population ordering in the system’s eigenbasis, defined as
\begin{equation}
    {C} = \langle \hat{H}^-_q \rangle - \langle \hat{H}^+_q \rangle,
\end{equation}
 where $\langle \hat{H}^-_q \rangle$ and $\langle \hat{H}^+_q \rangle$ denote the expectation values of the system Hamiltonian $\hat{H}_q$ evaluated with respect to the antipassive and passive states.  A finite capacity indicates the presence of non-passive population ordering that enables cyclic energy exchange, reflecting the system’s ability to accumulate and release energy coherently. Thus, the capacity quantifies how much energy the battery can store and supply through its population ordering in the energy eigenbasis.

The other conventionally defined thermodynamic quatities related to energetics in working battery  are the steadystate \textit{flux}, \textit{work done}, and \textit{power output} in the steady state. The output flux in the system is determined by the population transfer rate between levels of the work-mode and describes the rate at which energy is exchanged. We define the flux as the net transfer rate between states \( |\alpha\rangle \) and \( |\beta\rangle \):  
\begin{equation}
    j = e \Gamma_{\alpha\beta} \rho_{\alpha\alpha}.
    \label{eq:flux}
\end{equation}  
The  population imbalance between \( |\alpha\rangle \) and \( |\beta\rangle \),  dictates energy transfer into the work mode. The ratio of populations is given by $ 
    \zeta = \rho_{\alpha\alpha}/\rho_{\beta\beta},
   $   
the logarithm of which  defines the thermodynamic force associated with the energy exchange. The work done per transition is then given by  
\begin{equation}
    \mathcal{W}_{\alpha\beta} = (E_{\alpha} - E_{\beta}) + k_b T_w \ln \zeta,
    \label{eq:work}
\end{equation}
which  accounts for both the direct energy difference between levels and an additional entropic contribution due to the steady-state population imbalance.  The output power of the system, which characterizes the rate of useful energy transfer, can be quantified by defining 
\begin{equation}
    \mathcal{P} = j\, \mathcal{W}_{\alpha\beta}
    \label{eq:power}
\end{equation}
 In the regime where \( \zeta > 1 \), the battery behaves as a work supplier, converting energy from population transfer into useful work. Conversely, for \( \zeta \le 1 \), energy flows back into the system aiding the process of storage.

\section{RESULTS AND DISCUSSION}

\begin{figure}[h]
\centering
\includegraphics[width=1\linewidth]{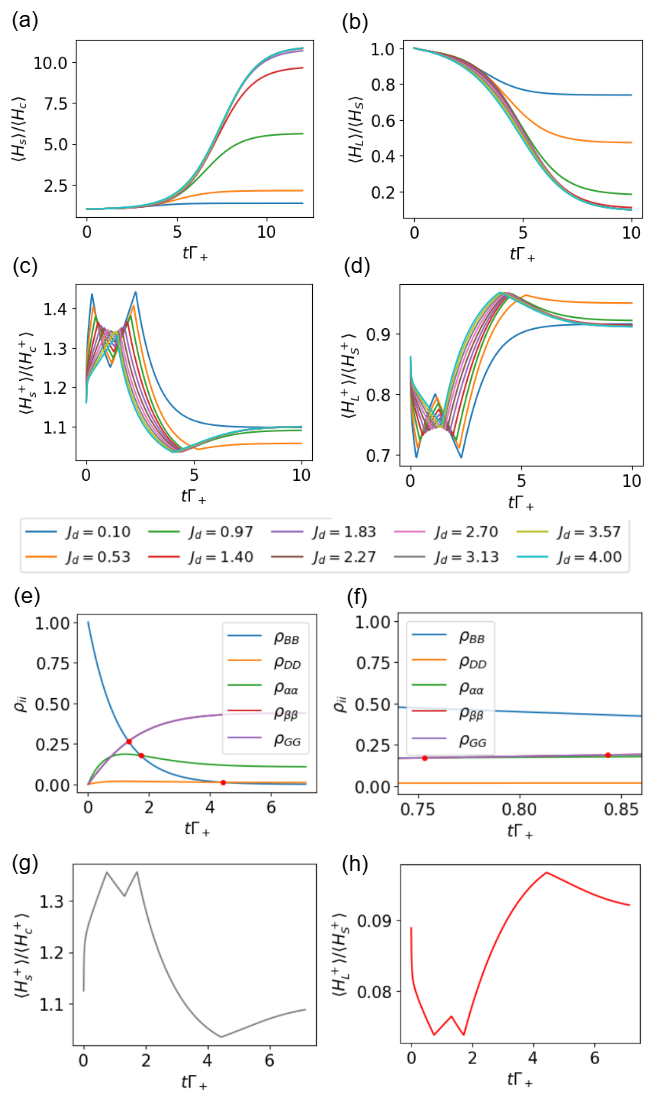}
\caption{ (a) Stored-to-charged energy, $\langle H_S\rangle / \langle H_C\rangle$ as a function of the battery's dimensionless time evolution ($t\Gamma_+$) and ring-levels and central system coupling strength $J_d$. 
(b) Leaked-to-stored energy, $\langle H_L\rangle / \langle H_S\rangle$. (c) Stored passive-to-charged passive energy, $\langle H_S^{+}\rangle / \langle H_C^{+}\rangle$. (d) Leaked passive-to-stored passive energy, $\langle H_L^{+}\rangle / \langle H_S^{+}\rangle$. (e-h) Same quantities as in (a-d) shown as contour plots vs. number of ring levels $N_R$ and $J_d$.
}
\label{fig-q_bat}
\end{figure}


 We numerically evaluate the density matrix $\rho$ by solving the quantum master equation and identify the passive states $\rho^+$ that emerge during the system's evolution, for a fixed configuration of the battery parameters. 
In Fig.~\ref{fig-q_bat}(a) we plot the time evolution of the storage-to-charging energy ratio, $\langle H_S\rangle/\langle H_C\rangle$, for different values of the coupling parameter $J_d$. Starting from a value close to unity, the ratio increases with time and exhibits a sigmoidal growth with the highest storage achieved at steadystate. The ratio  is always greater than unity proving that the storage is larger than the charging.  The long-time saturation value depends strongly on $J_d$. As $J_d$ is increased, the steady-state ratio becomes substantially larger than lower $J_D$ values, reaching values an order of magnitude above the initial level for the largest $J_d$ considered. This demonstrates that tuning $J_d$ (the coupling between the central system and the ring emitters) can markedly enhance the fraction of energy retained in the battery relative to the charging energy in the long-time regime.

In Fig.~\ref{fig-q_bat}(b) we show the corresponding leakage-to-storage energy ratio, $\langle H_L\rangle/\langle H_S\rangle$. In contrast to panel (a), this ratio decreases  from an initial value near unity and approaches a $J_d$-dependent steady-state plateau. The decay occurs in a reverse sigmoidal manner, with the most pronounced drop occurring at intermediate times, highlighting that the leakage is  less in comparison to storage. Importantly, the steady-state leakage fraction decreases strongly with increasing $J_d$. For smaller values of $J_d$, $\langle H_L\rangle/\langle H_S\rangle$ saturates at a relatively high value, whereas for larger $J_d$ the plateau is strongly suppressed, reaching values close to $\sim 0.1$. Taken together, panels (a) and (b) indicate that increasing $J_d$ simultaneously boosts the steady-state storage relative to charging while reducing the relative leakage, thereby improving the overall energy-retention performance of the battery.

In Fig.~\ref{fig-q_bat}(c) we plot the passive counterpart of the storage-to-charging ratio, $\langle H_S^{+}\rangle/\langle H_C^{+}\rangle$, which tracks the minimum (unitarily unextractable) energyof the state with time. Unlike panel (a), this ratio stays close to unity and shows an overshoot-dip-recovery in the form of kinks. This is because of the passive state changing discretely whenever the nonequilibrum populations cross. We show the population crossings in panels of Fig.~\ref{fig-q_bat}(e)-(f). The dynamics proceeds through six passive orderings. The resulting behavior of the ratio $\langle H_S^{+}\rangle/\langle H_C^{+}\rangle$ due to the cross-overs  is shown explicitly in  Fig.~\ref{fig-q_bat}(g) for a single representative value of $J_d$. For $0\lesssim t\Gamma_{+}<0.75$ the passive state is $\mathrm{diag}\{\rho_{BB},\rho_{\alpha\alpha},\rho_{\beta\beta},\rho_{GG},\rho_{DD}\}$,  for $0.75\lesssim t\Gamma_{+}<0.84$ the passive state becomes $\mathrm{diag}\{\rho_{BB},\rho_{\beta\beta},\rho_{\alpha\alpha},\rho_{GG},\rho_{DD}\}$, for $0.84\lesssim t\Gamma_{+}<1.33$ it further changes to $\mathrm{diag}\{\rho_{BB},\rho_{GG},\rho_{\beta\beta},\rho_{\alpha\alpha},\rho_{DD}\}$, for $1.33\lesssim t\Gamma_{+}<1.73$ the passive state switches to $\mathrm{diag}\{\rho_{\beta\beta},\rho_{GG},\rho_{BB},\rho_{\alpha\alpha},\rho_{DD}\}$, for $1.73\lesssim t\Gamma_{+}<4.44$ it becomes $\mathrm{diag}\{\rho_{\beta\beta},\rho_{GG},\rho_{\alpha\alpha},\rho_{BB},\rho_{DD}\}$ and for $t\Gamma_{+}\gtrsim4.44$ the system reaches the final passive state $\mathrm{diag}\{\rho_{\beta\beta},\rho_{GG},\rho_{\alpha\alpha},\rho_{DD},\rho_{BB}\}$.
 All these reordering produce the kinks features; once the final ordering is reached, the curve becomes smooth and saturates.

Figure~\ref{fig-q_bat}(d) shows the passive leakage-to-storage ratio, $\langle H_L^{+}\rangle/\langle H_S^{+}\rangle$, which exhibits a complementary transient (initial dip followed by a hump). The leakage and storage passive baselines respond differently to the same ordering switches. Panels (g)-(h) illustrate that the extrema align with the time windows where the passive state changes. Thermodynamically, panels (c)-(d) quantify how the system's passive energy is redistributed during relaxation, separating genuine retention effects [panels (a)-(b)] from spectral reordering effects that govern the ergotropy. With increasing $J_d$, the passive ratios in (c) and (d) show faster and smoother transients (weaker kinks and a quicker rise after the dip), while their long-time plateaus change only weakly compared to the strong $J_d$-dependence of the actual ratios in (a) and (b).

\begin{figure}[h]
\centering
\includegraphics[width=1\linewidth]{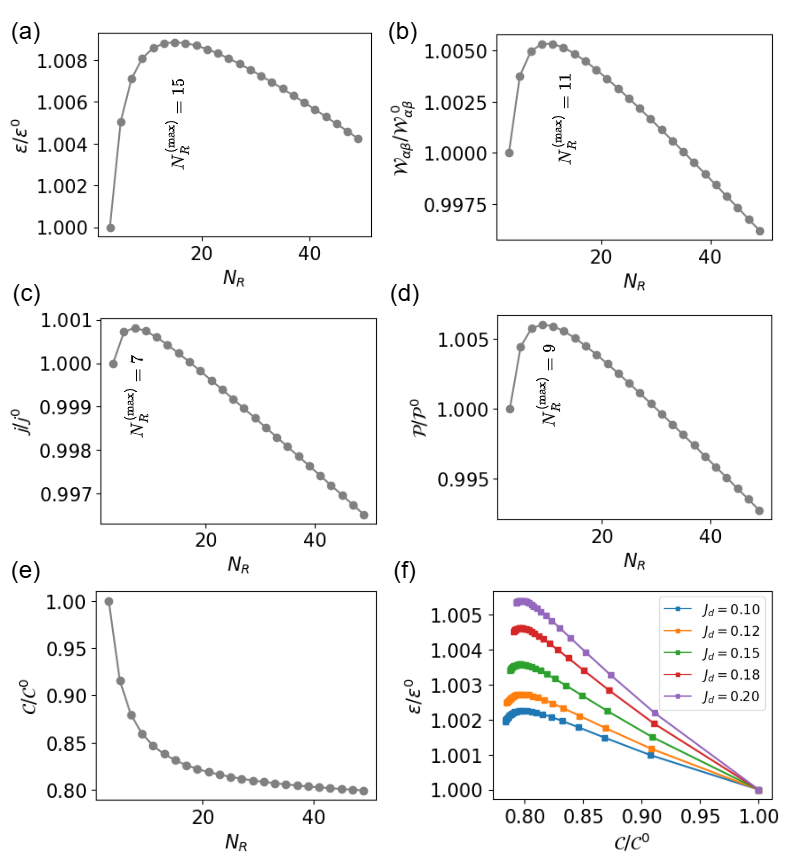}
\caption{ 
Variation of thermodynamic observables with the number of ring levels $N_R$ in the quantum battery model, (a) ergotropy, (b) work, (c) steady-state flux $j$, (d) output power and (e) capacity. (f) capacity Vs ergotropy with the number of ring levels $N_R$ and ring-levels and central system coupling strength $J_d$. To make each thermodynamic quantity dimensionless, these are plotted relative to the quantity's value with the smallest ring size $N_R =3$ and is denoted by a superscript, $^0$.
}

\label{fig-ther_quan}
\end{figure}

We now focus on the steadystate ergotropy, work, flux, power and capacity as a function of the number of ring levels, $N_R$ in Fig. (\ref{fig-ther_quan}a-e). We observe that ergotropy, flux, power, and work, each exhibit a distinct peak as a function of the number of ring levels. This reflects a separation between the optimal condition for storing extractable energy and that for transferring or converting it efficiently. Ergotropy peaks where the system achieves maximal deviation from passivity, Fig. (\ref{fig-ther_quan}a), which happens at $N_R  =15$ for the chosen numerical parameters. The total work, Fig. (\ref{fig-ther_quan}b) reaches its maximum at yet another optimal ring size ($N_R= 11$).  The flux, Fig. (\ref{fig-ther_quan}c) and power, Fig. (\ref{fig-ther_quan}d), also peak at different values of $N_R = 7,9$ respectively. 
The fact that each thermodynamic quantity attains a maximum at a distinct number of ring levels, reflect the different physical optimizations underlying each quantity. Ergotropy peaks when population inversion or deviation from the passive state is greatest, whereas flux is maximized at ring configurations that favor efficient energy transfer, balancing coupling and dynamic constraints. Power, being the instantaneous rate of work extraction, finds its optimum between these, where stored extractable energy and transfer dynamics are both favorable. The total work, which is the time integral of power, peaks at yet another ring size, representing the optimal trade-off between static energy resources and dynamic delivery over the operational interval. The difference in the optimal number is a direct consequence of the fact that both energy storage and transfer rates get aligned in different ways so as to optimize a different thermodynamic parameter. 

In Fig.~(\ref{fig-ther_quan}e) we show the normalized capacity $C/C^{0}$ as a function of the number of ring levels $N_R$. In contrast to ergotropy, work, flux, and power, the capacity is largest for the smallest rings and then decreases smoothly as $N_R$ increases, eventually approaching a nearly constant value for large
$N_R$. This behaviour means that the battery is most responsive for a ring  where only a few levels participate, the energy span between the antipassive and passive configurations is largest, whereas adding more ring levels spreads the excitation over a larger manifold,
reducing this energetic asymmetry and hence lowering the effective capacity.

\begin{figure}[h]
\centering
\includegraphics[width=1\linewidth]{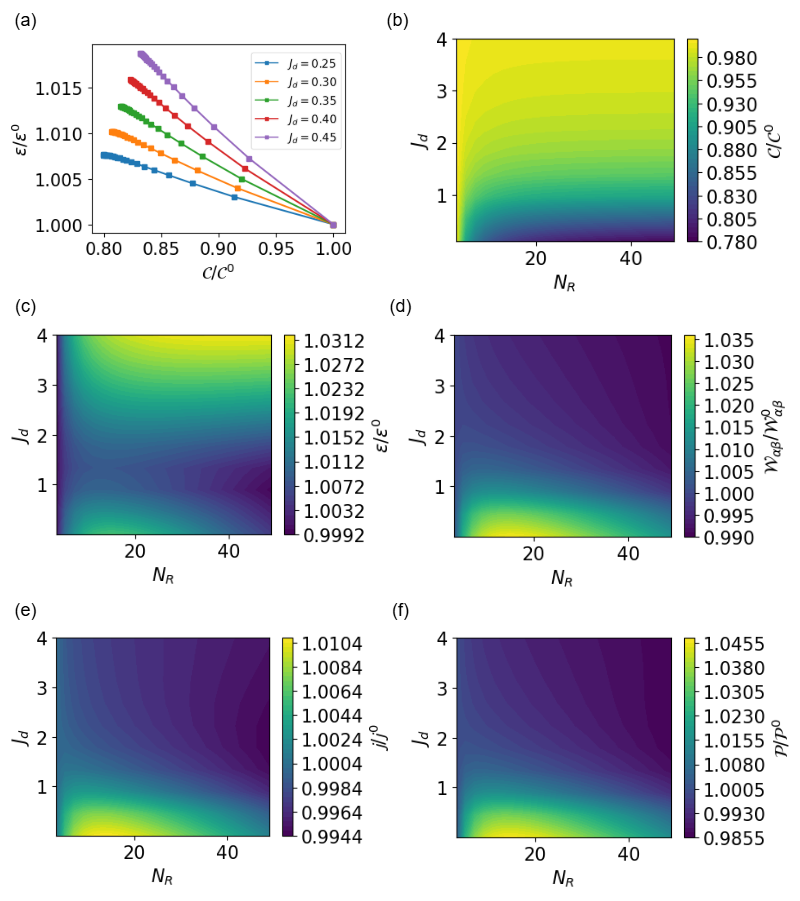}
\caption{Variation of thermodynamic observables with the number of ring levels $N_R$ and ring-levels and central system coupling strength $J_d$ in the quantum battery model, (a) capacity Vs ergotropy, (b) capacity (c) ergotropy, (d) work, (e) steady-state flux $j$ and (f) output power. To make each thermodynamic quantity dimensionless, these are plotted relative to the quantity's value with the smallest ring size $N_R =3$ and is denoted by a superscript, $^0$ at each value of $J_d$
}
\label{Jd_plot}
\end{figure}

Next we look at the inter-dependence between ergotropy and battery capacity. 
In Fig.(\ref{fig-ther_quan}f), we plot the two quantities, each evaluated as a function of increasing $N_R$. We see an optimal region at low values of $N_R$. Since the axis begins above unity, it indicates that the ergotropy always exceeds the capacity and is finally equal when $N_R$ is large.  As $N_R$ increases, both quantities decrease individually (Fig.\ref{fig-ther_quan}a and \ref{fig-ther_quan}b ) and and as a result of the decay to unity is interestingly linear. Increasing the coupling $J_d$ shifts this optimal region slightly towards higher values of $N_R$. For all values of $J_d$, the ratio eventually approaches unity. This is due to the fact that at large $N_R$ values, the contributions from anti-ergotropic (anti-passive) states become comparable to those of the active state. At still larger $J_d$, the optimal region  disappears entirely, as also evident in (Fig. \ref{Jd_plot}a). This indicates that strong coupling over-hybridizes the bright and dark
manifolds, suppressing the cooperative enhancement of ergotropy. Beyond this threshold, further increases in energetic polarization produce little or no additional extractable work. Thus, while ergotropy and
capacity are favorably correlated in the weak-coupling limit, the high-coupling regime (Fig.~\ref{Jd_plot}(a)) is dominated by geometric and dissipative constraints that flatten the energetic response of the quantum battery.
This behavior finds a natural analogue in the light‑harvesting complex 1 (LHC1) of purple bacteria \cite{swainsbury2023structure}. In LHC1, the ring size and geometric arrangement determines multiple figures of merit such as energy‑capture efficiency, coherence‑enhanced exciton transfer (supertransfer \cite{baghbanzadeh2016geometry}), and quinone diffusion pathways. Notably, structural flexibility in the LHC1 ring enables different functional optima. For example, closed rings enhance exciton delocalization and trapping efficiency, while slightly open or elliptical arrangements facilitate quinone traffic to the reaction center \cite{bahatyrova2004flexibility, swainsbury2023structure}. In a similar manner, our quantum battery’s performance metrics peak at different ring sizes, each optimized for its specific energetic function.

In Fig. (\ref{Jd_plot}b-f) we systematically investigate how the number of ring levels, 
$N_R$, influences the thermodynamic quantities for a range of ring-levels and central system coupling strength $J_d$. 
Figure~(\ref{Jd_plot}b) illustrates the contour map of the capacity $C/C^{0}$. The plot shows $C/C^{0}$ increases with $J_{d}$ and saturates along the higher coupling region, while decreasing smoothly with $N_{R}$. This behavior indicates that stronger coupling enhances reversible energetic polarization between the ergotropic and anti-ergotropic configurations, whereas larger ring sizes distribute the excitation more uniformly, leading to a weaker net polarization. The nearly flat contours at high $J_{d}$ demonstrate the saturation of capacity, suggesting that beyond this regime, further coupling does not significantly modify the energetic asymmetry. 
In Fig.~(\ref{Jd_plot}c), the steady-state ergotropy is intermediate at low values of $N_R$ and $J_d$, reaches a minimum at intermediate values of $N_R$ and $J_d$, and becomes maximum at higher values of $N_R$ and $J_d$. This is mainly due to the fact that the ergotropy is governed by two dominant population-imbalance contributions, $E_\alpha(\rho_\alpha-\rho_-)$ and $E_-(\rho_- - \rho_g)$. We interpret $E_\alpha(\rho_\alpha-\rho_-)$ as a storage contribution because $\ket{\alpha}$ and $\ket{-}$ belong to the storage manifold. We interpret $E_-(\rho_- - \rho_g)$ as a leakage contribution because $\ket{-}$ and the ground state $\ket{g}$ enter the leakage Hamiltonian. At low values of  $N_R$ and $J_d$, weak coupling allows only partial population transfer from the dark state $\ket{-}$ to the $\alpha$ state, creating a moderate imbalance between the contributing terms $E_\alpha(\rho_\alpha - \rho_{-})$ and $E_-(\rho_{-} - \rho_g)$ and yielding medium ergotropy. As both $N_R$ and $J_d$ increase to intermediate values, excessive hybridization of the dark and bright manifolds causes the storage term $E_\alpha(\rho_\alpha - \rho_-)$ and the leakage term $E_-(\rho_- - \rho_g)$ to nearly cancel, producing a minimum in ergotropy. For larger $N_R$ and stronger $J_d$, coherent coupling enhances population transfer from $\ket{-}$ to $\ket{\alpha}$, strengthening the storage contribution while the leakage channel saturates, thereby maximizing the ergotropy. Beyond this regime, ergotropy eventually saturates, signaling that over-coupling no longer yields proportional gains in extractable work. This saturation likely arises from excessive mixing between the ring and central system manifolds, which diminishes the population imbalance driving $E_\alpha(\rho_\alpha - \rho_-)$ and $E_-(\rho_- - \rho_g)$, thereby limiting further enhancement in ergotropy.

In contrast, panels (d)-(f) of Fig.~(\ref{Jd_plot}) show that work, flux, and power decrease with increasing coupling strength $J_d$. At weak to intermediate $J_d$, hybridization between the central two level system  and the collective ring mode promotes population transfer from the dark state $\lvert - \rangle$ to the $\alpha$-$\beta$ manifold, enhancing the excitation flux $j$ and thereby increasing both work $W_{\alpha\beta}$ and power $\mathcal{P}$. For larger $J_d$, however, excessive hybridization disrupts the spectral and spatial matching required for efficient ring-to-center transfer, effectively detuning the energy-transfer channel. This overcoupling suppresses the steady-state populations of $\lvert - \rangle$ and $\lvert \alpha \rangle$ through enhanced relaxation to the ground state, reducing the population imbalance $\zeta = \rho_{\alpha\alpha}/\rho_{\beta\beta}$ that drives work extraction and, consequently, diminishes energy exchange with the work reservoir. All three observables i.e $j/j^{0}$, $\mathcal{W}/\mathcal{W}^{0}$, and $\mathcal{P}/\mathcal{P}^{0}$ exhibit a nonmonotonic dependence on $N_{R}$ for each coupling strength. Each reaches an optimum where coherent transfer ($- \rightarrow \alpha \rightarrow \beta$) and dissipative loss ($\beta \rightarrow g$) are balanced. Beyond this point, stronger coupling induces population depletion through leakage, reducing flux, work, and power. The behavior is indicative of an interplay. Strong coupling enhances the system’s capacity to store energy (ergotropy), but diminishes the ability of delivering or converting that energy into useful work and power output.

From the preceding numerical results, we claim that the thermodynamic performance of the ring-central system based battery is governed by the interplay between ring size and coupling strength. Battery capacity increases with coupling and saturates at high $J_d$, while larger ring sizes alter the energetic asymmetry by distributing excitations more evenly. Steady-state ergotropy is set by the competition between the storage contribution $E_\alpha(\rho_\alpha - \rho_-)$ and the leakage term $E_-(\rho_- - \rho_g)$. At low $N_R$ and $J_d$, partial population transfer yields moderate ergotropy. At intermediate values, over-hybridization of bright and dark manifolds nearly cancels these contributions. At large $N_R$ and strong $J_d$, coherent coupling enhances population transfer to the storage manifold while leakage saturates, maximizing ergotropy before it plateaus. This could be  mainly due to excessive hybridization between the ring and center. In contrast, work, excitation flux, and power are nonmonotonic in $N_R$ and decline at strong coupling, as over-hybridization detunes the ring to center transfer channel and reduces the population imbalance driving energy extraction. It  emphasizes that factors controlling the population imbalance is fundamental to understand a quantum battery's characteristics.

\section*{Conclusion}
In this work, we proposed a complete working model of quantum battery that is inspired by photosynthetic light-harvesting system architecture. The model makes use of the explicit integration of cooperative couplings between geometrically arranged quantum systems and super and subradiant state engineering along with a cavity to mediate the storage subspace.  Through a numerical analysis based on an analytical master equation combined with effective analytical non-Hermitian modeling of the bright and dark states, we explored the ring-structured quantum battery's storage, leakage, and charging. By tracking the evolution of the system's reduced density matrix and identifying the sequence of passive states it traverses, we also explored the ergotropy and battery capacitance. We also introduced additional metrics, or passive counterparts,  such as the maximum possible charging,  the minimum available storage and the minimum unavoidable leakage. We elucidated the emergence of well-defined dynamical regimes characterized by abrupt transitions in these energy metrics due to the system transitioning between active and passive states.
Our results show that the storage-to-charging energy ratio increases sigmoidally from near unity, remains above unity at all times, and reaches a much larger steady-state value as the ring to center coupling is increased. It demonstrated enhanced long-time energy retention. In contrast, the leakage-to-storage ratio decreases from near unity in a reverse-sigmoidal manner and saturates at much smaller values for stronger coupling, indicating effective suppression of leakage relative to storage. The passive counterparts of these ratios remain close to unity but exhibit overshoot-dip-recovery and dip-then-hump transients with visible kinks, which arise from discrete passive-state reorderings induced by population crossings.  Notably, the dynamical evolution is punctuated by changes in the ordering of populations, which we analytically identified, capable of  directly mapping onto shifts in energy retention  and leakage suppression.

Upon systematically varying the number of ring levels, we found a nontrivial size dependence of battery performance. Ergotropy, total work, flux, and power each peak at different ring sizes which reiterate existing knowledge of the fact  that the optimal condition for storing extractable energy is not the same as that for transferring or delivering it efficiently. Ergotropy is maximized where deviation from passivity is greatest, which happens at a different value of ring size than the value at which the flux and power peak. The total work is optimized at an intermediate regime indicating that the storage and delivery need to act together favorably. The difference in the optimal number of ring moieties  is a direct consequence of the fact that both energy storage and transfer rates get aligned in different ways so as to optimize a particular thermodynamic parameter.   In contrast, the normalized capacity is largest for the smallest rings and decreases smoothly with increasing ring size before approaching an almost constant value for large rings, indicating that smaller manifolds support stronger energetic assymetry between passive and anti-passive configurations. This separation of optimal conditions mirrors design principles observed in biological systems such as the light-harvesting complexes, where structure-function relationships tune multiple energetic performance metrics independently.

We further examined the interdependent correlation between ergotropy and capacity both evaluated at different ring sizes.  The dependence of ergotropy on the capacity through the number of ring moieties shifts slightly toward higher ring size with increasing coupling between the ring and central system. This optimal region disappears at strong coupling  that  suppress cooperative enhancement and flattens the energetic response of the battery. Thus, ergotropy and capacity are favorably correlated in the weak-coupling regime, while the high-coupling regime is dominated by geometric and dissipative constraints.  Upon joint variation of the ring size and the ring to center coupling, we uncovered a compromise between storage enhancement and energy delivery. Stronger coupling increases the capacity and can maximize steady-state ergotropy. Whereas work, flux, and power decrease at large coupling. This was qualitatively attributed to the suppressing of the population imbalance needed for useful output. Through cooperative enhancement, thermodynamic energetics  are optimizable only within a finite parameter window where system variables that affect the cooperation are properly balanced. Overall, our results show that system geometry, passive-state structure, and collective coupling need to be  used as practical control knobs for optimizing different thermodynamic objectives in quantum batteries.
   

\bibliographystyle{unsrt}
\bibliography{references}

\end{document}